\documentclass[12pt]{article}%
\usepackage{amssymb}
\usepackage{amsmath}
\usepackage{amsfonts}
\usepackage{graphicx}%
\providecommand{\U}[1]{\protect\rule{.1in}{.1in}}

\begin{document}
\bigskip%
\begin{titlepage}
\vspace{.3cm} \vspace{1cm}
\begin{center}
\baselineskip=16pt \centerline{\bf{\Large{Asymptotically Free Mimetic Gravity}%
}}
\vspace{1truecm}
\centerline{\large\bf Ali H.
Chamseddine$^{1,2}$\ , Viatcheslav Mukhanov$^{3,4,5}$\ , Tobias B. Russ$^{3}%
$\ \ } \vspace{.5truecm}
\emph{\centerline{$^{1}%
$Physics Department, American University of Beirut, Lebanon}}
\emph{\centerline{$^{2}%
$Institute for Mathematics, Astrophysics and Particle Physics,}}
\emph{\centerline{Radboud University Nijmegen, Heyendaalseweg 135, 6525 AJ}}
\emph{\centerline{Nijmegen, The Netherlands}}
\emph{\centerline{$^{3}%
$Theoretical Physics, Ludwig Maxmillians University,Theresienstr. 37, 80333 Munich, Germany }%
}
\emph{\centerline{$^{4}%
$MPI for Physics, Foehringer Ring, 6, 80850, Munich, Germany}}
\emph{\centerline{$^{5}%
$School of Physics, Korea Institute for Advanced Study, Seoul 02455, Korea}}
\end{center}
\vspace{2cm}
\begin{center}
{\bf Abstract}
\end{center}
The idea of ``asymptotically free'' gravity is implemented using a constrained mimetic scalar field. The effective gravitational constant is assumed to vanish at some limiting curvature.
As a result singularities in spatially flat Friedmann and Kasner universes are avoided. Instead, the solutions in both cases approach a de Sitter metric with limiting curvature.
We show that quantum metric fluctuations vanish when this limiting curvature is approached.
\end{titlepage}%

\section{Introduction}

In \cite{mimetic} mimetic matter was introduced utilizing reparametrization of
the physical metric $g_{\mu\nu}$ in terms of an auxiliary metric $h_{\mu\nu}$
and a scalar field $\phi$ in the form
\begin{equation}
g_{\mu\nu}=h_{\mu\nu}h^{\alpha\beta}\phi_{,\alpha}\phi_{,\beta}\label{1}%
\end{equation}
This definition implies that $\phi$ identically satisfies
\begin{equation}
g^{\mu\nu}\phi_{,\mu}\phi_{,\nu}=1.\label{2}%
\end{equation}
Because the physical metric is invariant under Weyl transformations of
$h_{\mu\nu}$, the trace of the equations obtained by variation of the Einstein
action with respect to the metric vanishes identically. In the absence of matter
these equations become%
\begin{equation}
G_{\nu}^{\mu}-G\phi^{,\mu}\phi_{,\nu}=0,\label{3}%
\end{equation}
where $G_{\nu}^{\mu}=R_{\nu}^{\mu}-\frac{1}{2}\delta_{\nu}^{\mu}R$ is the
Einstein tensor, and they do not imply that $R=0$ even in vacuum. Therefore,
equation (\ref{3}) taken together with (\ref{2}) has additional solutions
imitating dust-like cold dark matter. The scalar field $\phi$ satisfies a
first order differential equation (\ref{2}) and hence has half a degree
of freedom which, when combined with the non-dynamical longitudinal mode of
gravity, provides an extra degree of freedom in the form of mimetic ``dust''.
Equivalently, the same theory is obtained by implementing equation (\ref{2}) as
a constraint added to the Einstein action:
\begin{equation}
S=\frac{1}{2}\int\textup{d}^{4}x\sqrt{-g}\left(  -\frac{1}{8\pi G}%
R+\lambda\left(  g^{\mu\nu}\phi_{,\mu}\phi_{,\nu}-1\right)  \right)
,\label{4}%
\end{equation}
where $\lambda$ is a Lagrange multiplier \cite{Golovnev}. Unexpectedly, the
concept of a mimetic field got a support in noncommutative geometry as a
consequence of the volume quantization of compact three dimensional foliations
of space time \cite{Quanta},\cite{Quanta2},\cite{Hilbert}. The mimetic field
$\phi$ proved to be very robust. It could be used to modify Einstein Gravity
in different possible ways. In particular, in \cite{mimcos} it was shown that
adding appropriate potentials $V\left(  \phi\right)  $ to the action leads to
many interesting cosmological solutions. Using instead gravity modification of
the Born-Infeld type, where $\Box\phi$ is bounded by a limiting value, allowed
to obtain bouncing solutions avoiding cosmological singularities
\cite{Singular} and to resolve black hole singularities \cite{BH}. Moreover, one
can use the mimetic field to easily construct ghost free massive gravity with
\textit{non} Fierz-Pauli mass term \cite{massivmim1},\cite{massivmim2}.

In this paper we will explore the possibility of a running gravitational
constant assuming that it depends on $\Box\phi,$ that is, $G=G\left(  \Box
\phi\right)  .$ As we shall see, this quantity is the only measure of
curvature $G$ can depend on without introducing higher time derivatives in the
modified Einstein equation. Assuming that $G$ vanishes at some limiting
curvature characterized by $\left(  \Box\phi\right)  _{L}^{2}$ we will
implement in this way the idea of ``asymptotic freedom'' for
gravity and investigate its possible consequences.

\section{Action and equations of motion}

Let us consider the theory with action%
\begin{equation}
S=\frac{1}{2}\int\textup{d}^{4}x\sqrt{-g}\left(  -f\left(  \Box\phi\right)
R-2\Lambda\left(  \Box\phi\right)  +\lambda\left(  g^{\mu\nu}\phi_{,\mu}%
\phi_{,\nu}-1\right)  +2\mathcal{L}_{m}\right)  , \label{4aa}%
\end{equation}
where%
\begin{equation}
f\left(  \Box\phi\right)  =\frac{1}{8\pi G\left(  \Box\phi\right)  }
\label{4bb}%
\end{equation}
is the inverse running gravitational constant, $\mathcal{L}_{m}$ is the matter
Lagrangian and for generality we also included a ``cosmological-like term'' $\Lambda\left(  \Box\phi\right)  \footnote{Please note
that we have changed the notations used in \cite{Singular} and \cite{BH} to
more convenient ones.}.$ Below we will use Planck units setting $8\pi G\left(
\Box\phi=0\right)  =8\pi G_0=1.$ In these units $f\left(  \Box\phi=0\right)
=1.$ Variation of the action with respect to the metric $g_{\mu\nu}$ gives
\begin{equation}
f G_{\mu\nu}+\left(  \,\Box f-\Lambda+\tfrac{1}{2}\left(  Z\,\phi^{;\alpha
}\right)  _{;\alpha}\right)  g_{\mu\nu}-f_{;\mu\nu}-Z_{(,\mu}\phi_{,\nu
)}=\lambda\phi_{,\mu}\phi_{,\nu}+T_{\mu\nu}^{(\mathrm{m}\textup{)}},
\label{5a}%
\end{equation}
where
\begin{equation}
Z:=Rf^{\prime}+2\Lambda^{\prime},
\end{equation}
$T_{\mu\nu}^{(\mathrm{m}\textup{)}}$ is the energy momentum tensor for matter
and the prime denotes derivative with respect to $\Box\phi.$ The equation%
\begin{equation}
\left(  Z^{;\nu}+2\lambda\phi^{;\nu}\right)  _{;\nu}=0 \label{7a}%
\end{equation}
follows from the variation of action (\ref{4aa}) with respect to $\phi$.
Alternatively (\ref{7a}) can be obtained as a consequence of the Bianchi
identities by taking the divergence of (\ref{5a}) and assuming that the energy
momentum tensor $T_{\mu\nu}^{(\mathrm{m}\textup{)}}$ for ordinary matter is
covariantly conserved. Taken together with the constraint
\[
g^{\mu\nu}\phi_{,\mu}\phi_{,\nu}=1,
\]
equation (\ref{7a}) allows to determine the Lagrange multiplier $\lambda.$

\section{The synchronous coordinate system}

The assumption of global solvability of (\ref{2}) is of course a restriction
on admissible space-times. As shown in \cite{HawkingEllis}, the existence of a
function whose gradient is everywhere time-like implies stable causality, i.e.
there are no closed time-like curves also for small perturbations of the
metric. Since the norm of the gradient of $\phi$ is not just positive but
everywhere equal to unity, $t:=\phi$ even qualifies to be used as the time
coordinate of a synchronous coordinate system (see \cite{Landau})

\begin{equation}
\textup{d}s^{2}=\textup{d}t^{2}-\gamma_{ik}\textup{d}x^{i}\textup{d}x^{k}
\label{5}%
\end{equation}
where the above equations greatly simplify.
In this coordinate system, the mimetic field $\phi$ defines the space-like
hypersurfaces of constant time. The extrinsic curvature of these
hypersurfaces,%
\begin{equation}
\kappa_{ik}=\frac{1}{2}\frac{\partial}{\partial t}\gamma_{ik} \label{7}%
\end{equation}
can be expressed as $\kappa_{ik}=-\phi_{;ik},$ while $\phi_{;0\alpha}=0.$
Thus,%
\begin{equation}
\Box\phi=g^{\alpha\beta}\phi_{;\alpha\beta}=\gamma^{ik}\kappa_{ik}%
=\kappa=\frac{\partial}{\partial t}\ln\sqrt{\gamma}, \label{8a}%
\end{equation}
that is, in this coordinate system $\Box\phi$ is simply equal to the trace of
the extrinsic curvature of the hypersurfaces of constant $\phi.$ In this
paper, for the sake of simplicity, we will only consider a homogeneous metric
with vanishing spatial curvature. In this case $\gamma_{ik}$ depends only on
time $t$ and equation (\ref{7a}) simplifies to%
\begin{equation}
\frac{1}{\sqrt{\gamma}}\partial_{t}\left[  \sqrt{\gamma}\left(  \partial
_{t}Z+2\lambda\right)  \right]  =0, \label{9a}%
\end{equation}
and can be easily integrated to give%
\begin{equation}
\lambda=-\frac{1}{2}\dot{Z}+\frac{C}{\sqrt{\gamma}} \label{10}%
\end{equation}
where the dot denotes derivative with respect to time $t$ and the constant of
integration $C$ describes the contribution of mimetic matter.

Substituting the expression (\ref{10}) for $\lambda$ in (\ref{5a}) and
calculating the covariant derivatives of $f$ and $Z$ we find that the $0-0$
component of the equation becomes%
\begin{equation}
fG_{00}+\left(  \dot{\kappa}+\frac{1}{2}R\right)  \kappa f^{\prime}%
-\Lambda+\kappa\Lambda^{\prime}=\varepsilon, \label{11}%
\end{equation}
where
\begin{equation}
\varepsilon \equiv T_{00}+\frac{C}{\sqrt{\gamma}}, \label{12}%
\end{equation}
is the total energy density of mimetic and ordinary matter.~Assuming that
the spatial components of the energy-momentum tensor satisfy $T_{k}%
^{i}\propto\delta_{k}^{i}$, subtracting from the spatial components of
equations (\ref{5a}) one third of their trace gives%
\begin{equation}
f\left(  G_{k}^{i}-\frac{1}{3}G_{m}^{m}\delta_{k}^{i}\right)  -\left(
f_{;k}^{;i}-\frac{1}{3}f_{;m}^{;m}\delta_{k}^{i}\right)  =0. \label{13}%
\end{equation}
For the spatially flat metric $\gamma_{ik}$
\begin{equation}
R_{0}^{0}=-\dot{\kappa}-\kappa_{k}^{i}\kappa_{i}^{k},\text{ \ }R_{k}%
^{i}=-\tfrac{1}{\sqrt{\gamma}}\partial_{0}\left(  \sqrt{\gamma}\kappa_{k}%
^{i}\right)  , \label{9}%
\end{equation}
where $\kappa_{k}^{i}=\gamma^{im}\kappa_{mk}$ (see, for example,
\cite{Landau}). Using these expression, equations (\ref{11}) and (\ref{13})
become
\begin{equation}
\frac{1}{3}\left(  f-2\kappa f^{\prime}\right)  \kappa^{2}-\Lambda
+\kappa\Lambda^{\prime}-\frac{1}{2}\left(  f+\kappa f^{\prime}\right)
\tilde{\kappa}_{k}^{i}\tilde{\kappa}_{i}^{k}=\varepsilon\label{14}%
\end{equation}
and
\begin{equation}
\partial_{0}\left(  f\sqrt{\gamma}\tilde{\kappa}_{k}^{i}\right)  =0,
\label{15}%
\end{equation}
correspondingly, where%
\begin{equation}
\tilde{\kappa}_{k}^{i}=\kappa_{k}^{i}-\frac{1}{3}\kappa\delta_{k}^{i},
\label{16}%
\end{equation}
is the traceless part of the extrinsic curvature.

The absence of higher time derivative terms in the modified Einstein
equations can be understood by realizing that in the synchronous coordinate
system
\begin{equation}
f R=f\left(  -2\dot{\kappa}-\kappa^{2}-\kappa_{k}^{i}\kappa
_{i}^{k}-{^{3}\!R}\right)  =-2\dot{F}-f\left(  \kappa
^{2}+\kappa_{k}^{i}\kappa_{i}^{k}+{^{3}\!R}\right)  \label{17}%
\end{equation}

where $f$ is assumed to be integrable with $f(\kappa)=F^{\prime}(\kappa)$ and
$^{3}\!R$ is the spatial curvature scalar. Hence the action contains, up to a
total derivative
only first order time derivatives of the metric.\footnote{This argument can, however,
only serve as a heuristic explanation. Strictly speaking, it is not
allowed to use $\Box\phi=\kappa$ and impose gauge conditions in the action
before variation.} This is a distinguishing feature of the $f(\Box\phi)$-theory
which would not be present if $\Box\phi$ is replaced by any other
non-constant, covariant expression containing first time derivatives of the
metric like e.g. $\phi^{;\mu\nu}\phi_{;\mu\nu}=\kappa_{k}^{i}\kappa
_{i}^{k}$.

Note that if we choose $f$ and $\Lambda$ to be symmetric functions, then
the time reversal invariance of the Einstein equation is maintained. Hence the
expanding counterparts for all the contracting solutions presented in the
following can be found simply by reversing the arrow of time.

\section{Asymptotic freedom and the fate of a collapsing universe}

Equation (\ref{14}) can be further simplified by making the choice
\begin{equation}
\Lambda=\frac{2}{3}\kappa^{2}(f-1) \label{18}%
\end{equation}
such that it becomes%
\begin{equation}
\left(  f-\frac{2}{3}\right)  \kappa^{2}-\frac{1}{2}\left(  f+\kappa
f^{\prime}\right)  \tilde{\kappa}_{k}^{i}\tilde{\kappa}_{i}^{k}=\varepsilon.
\label{19}%
\end{equation}
In our units the inverse gravitational constant $f$ is normalized to unity for
$\kappa^{2}=0.$ To guarantee that at low curvatures the corrections to General
Relativity will be in the next order in curvature we have to assume that for
$\kappa^{2}\ll1,$ $f=1+\mathcal{O}\left(  \kappa^{2}\right)  ;$ in this case
$\Lambda=\mathcal{O}\left(  \kappa^{4}\right)  .$ In addition we assume that the
gravitational constant $G(\kappa^{2})\propto1/f$ \ vanishes at some limiting
curvature $\kappa_{0}^{2}$ (cf. \cite{Markov},\cite{MukBran},\cite{Mukbransol}) and
thus take the simplest possible function for $f$, namely%
\begin{equation}
f=\frac{1}{1-\left(  \kappa^{2}/\kappa_{0}^{2}\right)  }, \label{20}%
\end{equation}
where $\kappa_{0}^{2}$ is a free parameter of the theory and can be taken well
below the Planckian value.

\textbf{Friedmann Universe.} First let us consider a flat contracting
Friedmann universe with the metric%
\begin{equation}
ds^{2}=dt^{2}-a^{2}\left(  t\right)  \delta_{ik}dx^{i}dx^{k}. \label{21}%
\end{equation}
In this case%
\begin{equation}
\kappa=3\frac{\dot{a}}{a} \label{22}%
\end{equation}
and $\tilde{\kappa}_{k}^{i}$ vanishes. Therefore, equation (\ref{16}) is
satisfied identically and equation (\ref{19}) can be rewritten as%
\begin{equation}
\frac{1}{3}\kappa^{2}\left(  \frac{1+2\left(  \kappa^{2}/\kappa_{0}%
^{2}\right)  }{1-\left(  \kappa^{2}/\kappa_{0}^{2}\right)  }\right)
=\varepsilon .\label{23}%
\end{equation}
Before writing the exact solution for equation (\ref{23}), we first consider some of its asymptotic limits. For $\kappa^{2}/\kappa_{0}^{2}\ll1$ it
reduces in the leading order to the usual Friedmann equation%
\begin{equation}
\left(  \frac{\dot{a}}{a}\right)  ^{2}=\frac{1}{3}\varepsilon.\label{24}%
\end{equation}
For a contracting universe dominated by matter with equation of state
$p=w\varepsilon$ it has the solution
\begin{equation}
a\propto t^{\frac{2}{3(1+w)}}, \label{25}%
\end{equation}
for large negative $t$. At the moment when the curvature approaches its
limiting value, the gravitational constant begins to decrease and for
${1-\left(  \kappa^{2}/\kappa_{0}^{2}\right)}$ $ \ll1 ,$ equation (\ref{23}) can
be approximated by
\begin{equation}
\kappa^{2}=\kappa_{0}^{2}\left(  1-\frac{\kappa_{0}^{2}}{\varepsilon
}+...\right)  . \label{26}%
\end{equation}
In a contracting universe the scale factor $a$ decreases, while the energy
density grows as $\varepsilon\propto a^{-3(1+w)}.$ Hence, the solution of equation
(\ref{26}) approaches the contracting flat de Sitter universe with constant
curvature where the scale factor decreases as
\begin{equation}
a\propto\exp\left(  -\frac{\kappa_{0}t}{3}\right)  \label{27}%
\end{equation}
for $\kappa_{0}t\gg1.$ The gravitational constant $G\propto f^{-1}$ vanishes as
$1/\varepsilon$ when $\varepsilon\rightarrow\infty$. The singularity is thus
avoided as a result of the asymptotic freedom of gravity irrespective of the
matter content of the universe.

For $\varepsilon\propto\gamma^{-\frac{1+w}{2}}$ the differential equation
(\ref{23}) can be integrated to obtain the exact implicit solution for
$\kappa\left(  t\right)  .$ In fact, differentiating the logarithm of equation
(\ref{23}) with respect to time and taking into account that $\partial
\ln\gamma/\partial t=2\kappa$, we obtain a first order differential equation
which can be easily integrated to give
\begin{equation}
\frac{1+w}{2}\kappa_{0}t=\frac{\kappa_{0}}{\kappa}-\operatorname{atanh}%
\frac{\kappa}{\kappa_{0}}-\sqrt{2}\arctan\left(  \sqrt{2}\frac{\kappa}%
{\kappa_{0}}\right).  \label{27a}%
\end{equation}
One can easily verify that the asymptotics (\ref{25}) and (\ref{27}) are
smoothly connected in this solution. In particular, for large negative $t$ the
universe contracts according to (\ref{25}). However, as it follows from
(\ref{27a}), $\kappa\left(  t=0\right)  \simeq-0.6\kappa_{0}$ instead of
blowing up as it would for solution (\ref{25}) and for large positive $t$
our solution approaches the de Sitter asymptotic (\ref{27}).

In conclusion, the singularity is replaced by a smooth transition to a de
Sitter metric. This qualitative behavior follows most naturally from our
theory, independent of a specific choice of $f$ and $\Lambda$. Note that the
modified Friedmann equation is in general just a relation of the form
$\kappa^{2}(\varepsilon)$. Demanding that this relation is smooth, one-to-one,
bounded and has bounded slope, as it is necessary to ensure limiting
curvature, the only remaining possibility is for $\kappa$ to approach its
constant limiting value as $\varepsilon$ tends to infinity.\textit{ }

\textbf{Kasner Universe. }We now consider a contracting anisotropic Kasner
universe to find out what happens when the curvature approaches its limiting
value for which the gravitational constant vanishes. To simplify the formulae
we will set the energy density of matter to zero although all our conclusions
survive also in the presence of the matter. For an anisotropic universe
\begin{equation}
\gamma_{ik}=\gamma_{(i)}\left(  t\right)  \delta_{ik} \label{28}%
\end{equation}
and $\gamma=\gamma_{\left(  1\right)  }\gamma_{\left(  2\right)  }%
\gamma_{\left(  3\right)  }.$ The traceless part of the extrinsic curvature in
this case is nonvanishing and is determined by intergrating equation
(\ref{16}):%
\begin{equation}
\tilde{\kappa}_{k}^{i}=\frac{\lambda_{k}^{i}}{f\sqrt{\gamma}}, \label{29}%
\end{equation}
where $\lambda_{k}^{i}$ are constants of integration satisfying
$\lambda_{i}^{i}=0.$ Substituting this expression in equation (\ref{19}) and
using (\ref{20}) we obtain%
\begin{equation}
\frac{1}{3}\kappa^{2}\left(  \frac{1+2\left(  \kappa^{2}/\kappa_{0}%
^{2}\right)  }{1-\left(  \kappa^{2}/\kappa_{0}^{2}\right)  }\right)  =\frac
{1}{2}\frac{\left(  1+\left(  \kappa^{2}/\kappa_{0}^{2}\right)  \right)
\bar{\lambda}^{2}}{\gamma}, \label{30}%
\end{equation}
where $\bar{\lambda}^{2}=\lambda_{k}^{i}\lambda_{i}^{k}.$ Because $\kappa
=\dot{\gamma}/2\gamma$, this equation allows us to determine how the
determinant of the metric depends on time. Knowing $\gamma\left(  t\right)  $,
the components of the metric can be found in the following way: Without loss
of generality we can diagonalize $\lambda_{k}^{i},$ so that, $\lambda_{k}%
^{i}=\lambda_{(i)}\delta_{k}^{i}.$ Taking into account the definitions
(\ref{7}) and (\ref{16}), equations (\ref{29}) reduce to%
\begin{equation}
\frac{\dot{\gamma}_{(i)}}{\gamma_{(i)}}-\frac{1}{3}\frac{\dot{\gamma}}{\gamma
}=\frac{2\lambda_{(i)}}{f\sqrt{\gamma}}, \label{31}%
\end{equation}
from which it follows that%
\begin{equation}
\gamma_{(i)}=\gamma^{1/3}\exp\left(  \int\frac{2\lambda_{(i)}}{f\sqrt{\gamma}%
}dt\right).  \label{32}%
\end{equation}
Before giving the exact solution of equation (\ref{30}) it is more
enlightening to study the asymptotic solutions. At low curvatures, that is,
for $\kappa^{2}\ll\kappa_{0}^{2},$ equation (\ref{30}) simplifies to%
\begin{equation}
\left(  \frac{\dot{\gamma}}{\gamma}\right)  ^{2}\simeq\frac{6\bar{\lambda}%
^{2}}{\gamma}, \label{33}%
\end{equation}
and has the solution%
\begin{equation}
\gamma=\frac{3}{2}\bar{\lambda}^{2}t^{2}. \label{34}%
\end{equation}
Taking into account that in this limit $f=1$ and substituting this solution in
(\ref{32}) we find%
\begin{equation}
\gamma_{(i)}=\left(  \frac{3}{2}\bar{\lambda}^{2}\right)  ^{1/3}t^{2p_{i}},
\label{35}%
\end{equation}
where
\begin{equation}
p_{i}=\frac{1}{3}\pm\sqrt{\frac{2}{3}}\frac{\lambda_{(i)}}{\bar{\lambda}}.
\label{36}%
\end{equation}
Since $\lambda_{1}+\lambda_{2}+\lambda_{3}=0$, the $p_{i}$ satisfy the
conditions
\[
p_{1}+p_{2}+p_{3}=1,\text{ \ \ \ \ }p_{1}^{2}+p_{2}^{2}+p_{3}^{2}=1,
\]
and at low curvatures we have either an expanding or a contracting Kasner universe
\cite{Landau}.

In a contracting universe, at $\left\vert t\right\vert \simeq1/\kappa_{0}$ the
curvature becomes of the order of limiting curvature and for $1-\left(
\kappa^{2}/\kappa_{0}^{2}\right)  \ll1,$ equation (\ref{30}) is well
approximated by%
\begin{equation}
\kappa_{0}^{2}\left(  \frac{1}{1-\left(  \kappa^{2}/\kappa_{0}^{2}\right)
}\right)  =\frac{\bar{\lambda}^{2}}{\gamma}, \label{37}%
\end{equation}
from which it follows that
\begin{equation}
\frac{\dot{\gamma}}{\gamma}=-2\kappa_{0}\left(  1-\frac{\kappa_{0}^{2}\gamma
}{\bar{\lambda}^{2}}\right)  ^{1/2} \label{38}%
\end{equation}
in a contracting universe and for $\gamma\ll\bar{\lambda}^{2}/\kappa_{0}^{2}$
we have
\begin{equation}
\gamma\propto\exp\left(  -2\kappa_{0}t\right) . \label{39}%
\end{equation}
As follows from (\ref{38}), in this limit
\begin{equation}
f=\frac{\bar{\lambda}^{2}}{\kappa_{0}^{2}\gamma} \label{40}%
\end{equation}
and the integrals in (\ref{32}) fast converge to some constants for $t\gg
1/\kappa_{0}.$ These constants can be absorbed by redefinition of the spatial
coordinates to give the asymptotic solution%
\begin{equation}
\gamma_{\left(  1\right)  }=\gamma_{\left(  2\right)  }=\gamma_{\left(
3\right)  }=\gamma^{1/3}\propto\exp\left(  -\frac{2}{3}\kappa_{0}t\right)  ,
\label{41}%
\end{equation}
that describes a contracting flat de Sitter universe with constant curvature.

The exact implicit solution of equation (\ref{30}) for $\kappa\left(
t\right)  $ is given by
\begin{equation}
\kappa_{0}t=\frac{\kappa_{0}}{\kappa}-\operatorname{atanh}\frac{\kappa}%
{\kappa_{0}}-\sqrt{2}\arctan\left(  \sqrt{2}\frac{\kappa}{\kappa_{0}}\right)
+\arctan\frac{\kappa}{\kappa_{0}}.%
\end{equation}

Note that in the anisotropic case we are forced to use asymptotic freedom if
we want to obtain a non-singular modification where $\kappa$ tends to its
constant limiting value. Only in this way the anisotropy can disappear during contraction.

\section{Quantum fluctuations}

Now we look at what happens with quantum fluctuations of the gravitational
field as we approach the limiting curvature where the gravitational constant
vanishes. As it is well known (see, for example, \cite{Mbook}), in General
Relativity the typical amplitude of quantum fluctuations of gravitational
waves in Minkowski space and on curved background at scales $l\,$ much
smaller than the curvature scale is about
\begin{equation}
\delta h_{l}\simeq\frac{\sqrt{G}}{l}, \label{42}%
\end{equation}
where $G$ is the gravitational constant. Therefore, in our theory where this
gravitational constant vanishes on the background with limiting curvature, one
could expect that the quantum metric fluctuations must also vanish. We will
now show that this is what really happens. Consider a slightly perturbed flat
Friedmann Universe with metric%
\begin{equation}
ds^{2}=a^{2}\left(  \eta\right)  \left(  d\eta^{2}-\left(  \delta_{ik}%
+h_{ik}\right)  dx^{i}dx^{k}\right)  , \label{43}%
\end{equation}
where we have introduced conformal time $\eta=\int\frac{dt}{a\left(  t\right)
}$ and $h_{ik}$ is the traceless ($h_{i}^{i}=0$) and transverse ($h_{k,i}%
^{i}=0$) part of the metric perturbations. Substituting this metric in action
(\ref{4aa}) and expanding it to second order in $h$ we obtain the following
action for the gravitational waves:%
\begin{equation}
S=\frac{1}{8}\int fa^{2}\left(  h_{k}^{i\prime}h_{i}^{k\prime}-h_{k,m}%
^{i}h_{i}^{k,m}\right)  d\eta d^{3}x, \label{44}%
\end{equation}
where prime denotes the derivative with respect to conformal time $\eta$ and
the spatial indices are raised and lowered with $\delta_{ik}.$ This precisely
coincides with the action for gravitational waves in a Friedmann universe
with the ``scale factor''%
\[
\tilde{a}:=a\sqrt{f}.
\]
In this case the quantization procedure is well known and there is no need to
repeat all the steps here. Referring to section 8.4 in \cite{Mbook} we find
that the typical amplitude squared for the quantum fluctuations is%
\begin{equation}
\delta h^{2}\left(  k,\eta\right)  \simeq\frac{\left\vert v_{k}\right\vert
^{2}k^{3}}{\tilde{a}^{2}}=\frac{\left\vert v_{k}\right\vert ^{2}k^{3}}{fa^{2}%
}, \label{45}%
\end{equation}
where $k$ is the co-moving wave number and the mode function $v_{k}$ satisfies
the equation%
\begin{equation}
v_{k}^{\prime\prime}+\omega_{k}^{2}v_{k}=0,\text{ \ \qquad}\omega_{k}%
^{2}\equiv k^{2}-\frac{\tilde{a}^{\prime\prime}}{\tilde{a}}\text{\ \ }
\label{46}%
\end{equation}
with initial conditions $v_{k}\left(  \eta_{in}\right)  =1/\sqrt{\omega_{k}},$
$v_{k}^{\prime}\left(  \eta_{in}\right)  =i\sqrt{\omega_{k}}$ for quantum
fluctuations. When the solution approaches the limiting curvature we have
$f\propto\varepsilon\propto a^{-3(1+w)}$ and $\tilde{a}\propto a^{-\frac{1}%
{2}(1+3w)}.$ Taking into account that in contracting de Sitter $a\left(
\eta\right)  =3/\kappa_{0}\eta,$ where $\eta$ grows, equation (\ref{46})
becomes%
\begin{equation}
v_{k}^{\prime\prime}+\left(  k^{2}-\frac{\left(  9w^{2}-1\right)  }{4\eta^{2}%
}\right)  v_{k}=0 .\label{47}%
\end{equation}
We can define quantum fluctuations only for short wave gravitational waves
satisfying $k\eta\gg1,$ that is, for physical scales $l=a/k\ll\kappa_{0}%
^{-1}.$ In this case $v_{k}\simeq\exp\left(  ik\eta\right)  /\sqrt{k}$ and, as
follows from (\ref{45})%
\begin{equation}
\delta h\left(  l\right)  \simeq\frac{1}{\sqrt{f}l}\simeq\frac{\sqrt{G\left(
\kappa\right)  }}{l}. \label{48}%
\end{equation}
Hence, quantum fluctuations in a given physical scale $l\ll\kappa_{0}^{-1}$
vanish as $\kappa\rightarrow\kappa_{0}$ and correspondingly $G\left(
\kappa\right)  \rightarrow0.$ This is in complete agreement with our
expectations. The perturbations with $k\eta\ll1,$ which were outside the horizon
$\kappa_{0}^{-1}$ finally come inside because $\eta$ grows in a contracting de
Sitter space-time. The amplitude of metric perturbations $h$ is constant
before horizon crossing, but after entering the horizon it decays as
$\tilde{a}^{-1}\propto a^{\frac{1}{2}(1+3w)}.$ Thus we have shown that the de
Sitter space-time with limiting curvature is completely classical, with no
quantum metric fluctuations present.

\section{Conclusions}

The simple observation that the conformal part of the metric in General
Relativity can be extracted covariantly via a constrained scalar field $\phi$
has proven to be very fruitful. The resulting modified gravity theory
does not induce any additional degrees of freedom for the graviton, but at the
same time makes the longitudinal mode  dynamical even in the absence of matter.
This mode can serve as a viable candidate for dark matter in our
Universe. Moreover the constrained scalar field allows us to build 
invariants which in synchronous coordinates can be expressed
exclusively in terms of first order time derivatives of the metric. This opens
the possibility to modify General Relativity in a simple way avoiding
problematic higher order time derivative terms which generically lead to ghost
degrees of freedom. Such a generalization of Einstein theory happens to be very
interesting and allows us for example to implement the idea of limiting
curvature and resolve spacelike singularities in Friedmann and Kasner universes
as well as in black holes. The limiting curvature, which is a parameter of
the theory, can be taken well below the Planckian curvature. Potentially, this would make the difficult unresolved problem of non-perturbative quantum gravity obsolete for all practical purposes. 

In this paper we have investigated the possibility of implementing the idea
of classical asymptotic freedom just assuming that the gravitational constant
vanishes at the limiting curvature. As it was shown, in this case the
singularities in flat contracting Friedmann and Kasner universes are
resolved and close to the limiting curvature the de Sitter solution is approached.
Moreover, quantum metric fluctuations asymptotically vanish and the
spacetime becomes fully classical at this limiting curvature. This opens an
interesting possibility to resolve the longstanding singularity problem in
General Relativity via a simple modification of Einstein theory at large
curvatures without referring this problem to a yet unknown non-perturbative theory of
quantum gravity. 

For the sake of simplicity and to highlight the most important aspects first, in this paper we focused mainly on the homogeneous, spatially flat sector of the theory proposed above. In another soon to appear paper we will extend our analysis and consider applications to spatially non-flat spacetimes, including Black Holes.

\bigskip

\bigskip

\textbf{{\large {Acknowledgments}}}

The work of A. H. C is supported in part by the National Science Foundation
Grant No. Phys-1518371. The work of V.M. and T.B.R. is supported by the Deutsche
Forschungsgemeinschaft (DFG, German Research Foundation) under Germany's
Excellence Strategy -- EXC-2111 -- 390814868. V.M. is grateful to Korea
Institute for Advanced Study, where the part of this work was completed, for hospitality.


\begin{thebibliography}{99}                                                                                               %


\bibitem {mimetic}A. H. Chamseddine and V. Mukhanov, \textit{Mimetic Dark
Matter, }JHEP \textbf{1311 }(2013) 135.

\bibitem {Golovnev}A. Golovnev, \textit{On the Recently Proposed Mimetic Dark
Matter, }Phys. Lett. \textbf{B728 }(2014) 39.

\bibitem {Quanta}A. H. Chamseddine, A. Connes and V. Mukhanov, \textit{Quanta
of Geometry: Noncommutative Aspects, }Phys. Rev. Lett. \textbf{114 }(2015) 091302.

\bibitem {Quanta2}A. H. Chamseddine, A. Connes and V. Mukhanov,
\textit{Geometry and the Quantum: Basics, JHEP \textbf{1412} (2014) 098. }

\bibitem {Hilbert}A. H. Chamseddine \textit{Quanta of Space-Time and
Axiomatization of Physics, }in \textrm{Foundations of Mathematics and Physics
One Century After Hilbert, }pages 211-251, Editor J. Kouneiher, Springer 2018.

\bibitem {mimcos}A. H. Chamseddine, V. Mukhanov and A. Vikman,
\textit{Cosmology with Mimetic Matter, }JCAP \textbf{1406 }(2014) 017.

\bibitem {Singular}A. H. Chamseddine, V. Mukhanov, \textit{Resolving
Cosmological Singularities, }JCAP \textbf{1703 }(2017) 009.

\bibitem {BH}A. H. Chamseddine, V. Mukhanov, \textit{Nonsingular Black Hole,
}Eur. Phys. J. \textbf{C77} (2017) 83..

\bibitem {massivmim1}A. H. Chamseddine, V. Mukhanov, \textit{Ghost Free
Mimetic Massive Gravity, }JHEP \textbf{1806 }(2018) 060.

\bibitem {massivmim2}A. H. Chamseddine, V. Mukhanov, \textit{Mimetic Massive
Gravity: Beyond Linear Approximation, }JHEP \textbf{1806 }(2018) 062.

\bibitem {Markov}M. Markov, Pis'ma Zh. Eskp. Teor. Fiz. \textbf{36 }(1982)
214; \textbf{46 }(1987) 341 [JETP Lett. \textbf{36 }(1982) 265; \textbf{46
}(1987) 431.

\bibitem {MukBran}V. Mukhanov and R. Brandenberger, \textit{A Nonsingular
Universe, }Phys. Rev. Lett. \textbf{68 (}1992) 1969.

\bibitem {Mukbransol}V. Mukhanov, R. Brandenberger, and A. Sornborger,
\textit{A Cosmological Theory without Singularities, }Phys. Rev. D \textbf{48
}(1993) 1629.

\bibitem {Landau}L. Landau and E. Lifshitz, \textit{The Classical Theory of
Fields} fourth edition, Butterworth, Heinemann, 1980.

\bibitem {HawkingEllis}S. Hawking and G. Ellis, \textit{The Large Scale
Structure of Space-Time, }Cambridge University Press, 1975.

\bibitem {Mbook}V. Mukhanov, \textit{Physical Foundations of Cosmology,
}Cambridge University Press, 2005.
\end{thebibliography}
\end{document}